\documentclass[letterpaper,10pt]{article} 

\usepackage{graphicx}
\usepackage[font=bf,skip=\baselineskip]{caption}
\usepackage{subcaption}
\usepackage{booktabs}
\usepackage{amsmath}
\usepackage{algorithm}
\usepackage[noend]{algpseudocode}
\usepackage[usenames,dvipsnames,svgnames,table]{xcolor}
\usepackage[backend=bibtex,sorting=none,maxnames=2]{biblatex}

\graphicspath{{./}}
\definecolor{light-gray}{gray}{0.8}

\newcommand{\wv}{\texttt{word2vec}}
\newcommand{\dv}{\texttt{doc2vec}}
\newcommand{\tSNE}{\texttt{t-SNE}}

\def\keywords#1{\textbf{Keywords}:\emph{#1}}

\newcommand{\affaddr}{\normalsize\itshape}
\newcommand{\email}{\normalsize\tt}

\begin{document} 

\title{Gendered Conversation in a Social Game-Streaming Platform}
\author{Supun Nakandala$^{*}$, Giovanni Luca Ciampaglia$^{\dag}$, \and Norman 
Makoto Su$^{\ddag}$, and Yong-Yeol Ahn$^{\dag\ddag}$\\
\affaddr{$*$ Pervasive Technologies Institute}\\
\affaddr{$\dag$ Indiana University Network Science Institute} \\
\affaddr{$\ddag$ School of Informatics and Computing} \\
\affaddr{Indiana University, Bloomington, IN USA}\\
\email{\{snakanda, gciampag, normsu, yyahn\}@iu.edu}}

\date{}

\maketitle 

\begin{abstract} 

Online social media and games are increasingly replacing offline social
activities. Social media is now an indispensable mode of communication; online
gaming is not only a genuine social activity but also a popular spectator
sport. With support for anonymity and larger audiences, online interaction
shrinks social and geographical barriers.  Despite such benefits, social
disparities such as gender inequality persist in online social media. In
particular, online gaming communities have been criticized for persistent
gender disparities and objectification. As gaming evolves into a social
platform, persistence of gender disparity is a pressing question. Yet, there
are few large-scale, systematic studies of gender inequality and
objectification in social gaming platforms.  Here we analyze more than one
billion chat messages from Twitch, a social game-streaming platform, to study
how the gender of streamers is associated with the nature of conversation.
Using a combination of computational text analysis methods, we show that
gendered conversation and objectification is prevalent in chats. Female
streamers receive significantly more objectifying comments while male streamers
receive more game-related comments.  This difference is more pronounced for
popular streamers. There also exists a large number of users who post only on
female or male streams.  Employing a neural vector-space embedding (paragraph
vector) method, we analyze gendered chat messages and create prediction models
that (i) identify the gender of streamers based on messages posted in the
channel and (ii) identify the gender a viewer prefers to watch based on their
chat messages.  Our findings suggest that disparities in social game-streaming
platforms is a nuanced phenomenon that involves the gender of streamers as well
as those who produce gendered and game-related conversation. 

\end{abstract} 


\section{Introduction}\label{introduction} 

Simone de Beauvoir said, ``One is not born a woman, but becomes
one''~\cite{beauvoir2012second}, highlighting the situation of women as not
free to make decisions about their life, but rather shackled by a society that
objectifies them, severely limiting their actions and opportunities. Since
Beauvoir's clarion call, researchers have examined the extent to which women
continue to be objectified in popular media such as television, movies, and
advertisements~\cite{gill2008empowerment}.  Such media continue to reinforce
women as objects under the ``gaze'' of men~\cite{holland2004male}.

The Internet and the Web enable complex forms of many-to-many social
interaction and make one's identity less conspicuous. On first glance, they
provide an ostensibly ``gender-neutral'' medium, offering new opportunities to
empower women.  However, studies suggest that inequality remains in online
spaces. For instance, in the popular microblogging platform Twitter, the
``glass-ceiling'' effect~\cite{nilizadeh2016twitter} and gender-biased user
dialogue~\cite{garcia2014gender,fulper2015misogynistic} are observed. Studies on image search
engines~\cite{kay2015unequal} and Wikipedia~\cite{hill2013wikipedia,
wagner2015mans, graells-garrido2015first} also demonstrate persistent gender
stereotypes and disparities. 


While data-driven research on online gender inequality has focused on several
popular online platforms, online gaming has received little attention, although
the advent of the Internet and of social media has transformed video games into
genuine social activities~\cite{kaytoue2012watch}. Video games are no longer
the purview of arcades and family rooms; they are social activities connecting
people across the world, and a widely broadcasted and watched medium. Numerous
online communities are devoted to discussing, watching, and playing video
games. Traditionally considered a ``boy's activity''~\cite{cassell2000barbie,
su2011virtual}, the culture of gaming communities has been accused of
misogyny~\cite{massanari2015gamergate}. Videos games themselves can be a medium
that glorifies the objectification of women~\cite{dill2007video,
burgess2007sex, paassen2016true}. The online space of video games provides no
respite from these inequities; ethnographic studies have observed that, when
female gamers have revealed their identities online, gamers cease speaking
about game-related topics and instead shift to the gamer herself and her
gender~\cite{su2011virtual, nardi2010my}. 

Yet, little work has systematically examined, on a large scale, the possibly
gendered nature of the next evolution of social media and online video
gaming: social video game-streaming platforms. On the most popular of these
platforms, Twitch, gamers can stream their gameplay and communicate with
viewers in real-time. To give an idea of its popularity, in 2015, Twitch had a
monthly average of 1.7 million broadcasters and half a million concurrent
viewers~\cite{twitch2015retro}. The 2013 world championship of ``League of
Legends'', a popular online game, was broadcast live on Twitch; the event was
watched by more people than the NBA finals~\cite{mccormick2013league}.

An interesting aspect of Twitch is that the success of e-sports has allowed
some video game professionals and personalities to garner a cult following
rivaling those of many celebrities. Central to the success of these
individuals are their channels. In Twitch, each game stream is called a
``channel'' and is run by an individual streamer, a group of streamers, an
organization, or a channel aggregator. Browsing the list of public channels,
one can not only find a wide variety of games --- League of Legends, Counter
Strike, and Hearthstone~\cite{twitch2015retro, twingetv2016} --- but creative
performance arts such as painting, music, and animation making. Many streamers
play, in addition to their main game, other games in their channels. 

\begin{figure}[t] 
  \begin{center}
      \includegraphics[width=0.5\textwidth]{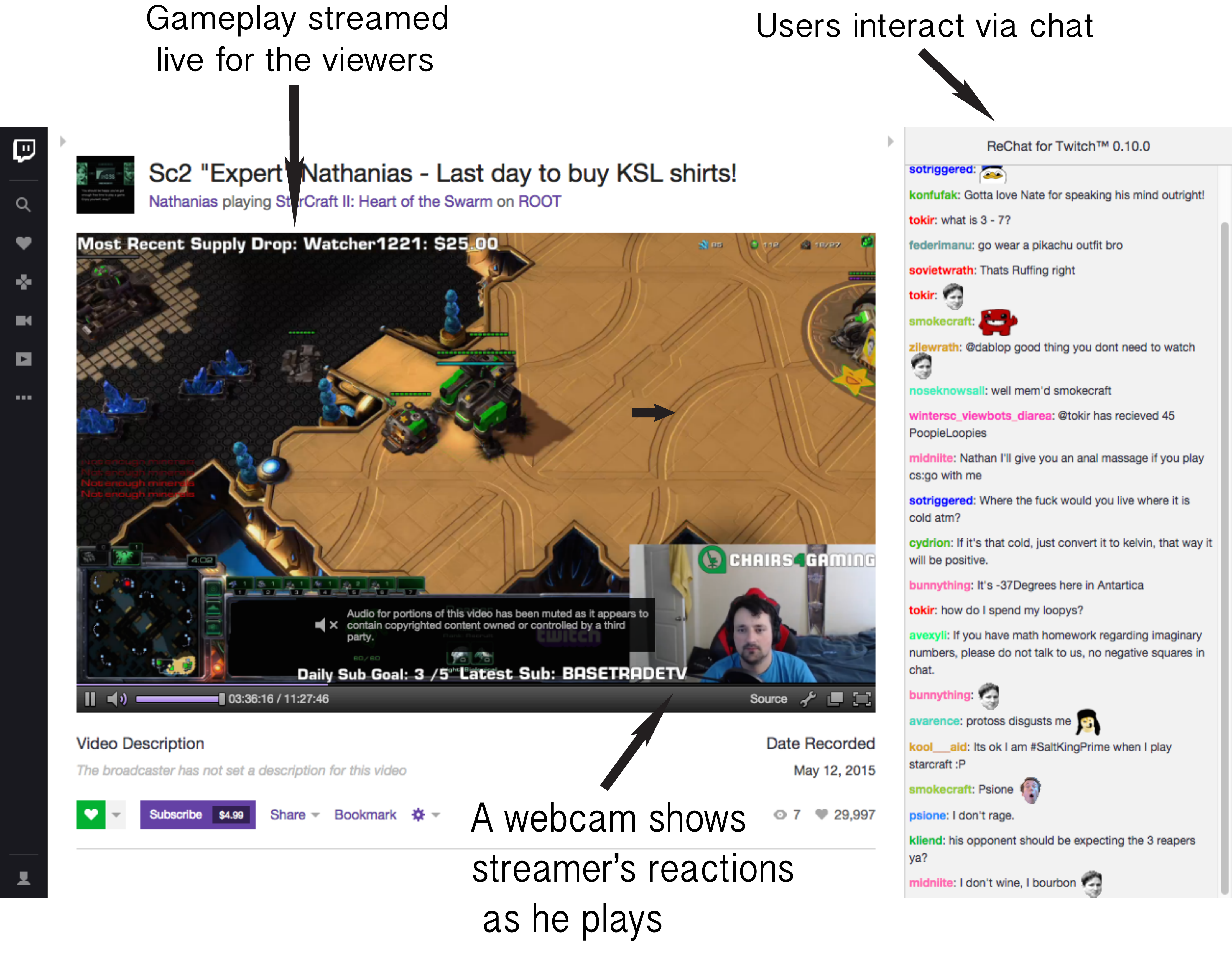}
  \end{center}

\caption{Twitch Channel Stream Interface}\label{fig:twitch_stream_img}

\end{figure}

Twitch facilitates communication between viewers and the streamers by providing
a public chat room for each channel. Viewers can post chat messages to
communicate with the streamers or with other viewers. Streamers can also post
in the chat, but often streamers just share their webcam feed and talk directly
in the stream. Importantly, the relationship between viewers and streamers is a
potential source of income in Twitch; streamers earn revenue by holding game
events, having subscribers, and encouraging donations.
Fig.~\ref{fig:twitch_stream_img} shows the interface of Twitch which has a game
feed, streamer's webcam feed, and public chat room. 

In sum, Twitch represents one of the most popular platforms for the rapidly
rising form of social gaming. In Twitch, the spectacle of gaming is not only
watching the video game and streamers but also the actions of the viewers
themselves~\cite{dalsgaard2008performing}. With this fundamental transformation
in gaming culture, where viewers and streamers both have the power to
communicate and to be seen, our study investigate how gender inequality
manifests in the Twitch platform, asking the following research questions:
\begin{itemize}
\item \emph{Are chat messages addressed to streamers gendered?} Is there a
relationship between the gender of a streamer and the nature of messages that
the streamer receives? For instance, do female streamers receive more
objectifying comments, while male streamers receive more game-related messages?
Is it possible to classify the gender of a channel's streamer by the comments
they receive?
\item \emph{Are viewers and their messages gendered?} Do viewers choose
channels based on gender? Is the gendered choice of channels correlated with
objectifying language? 
\end{itemize} 
Our analysis on whether social game-streaming platforms exhibit gendered
behavior is timely. These platforms have become a powerful and influential
medium for new and young gamers alike, and this influence may have far-reaching
consequences outside the domain of social media, for example by distorting
beliefs about women in the real world~\cite{behm-morawitz2009effects}.

\subsection{Ethics Statement}

This study was reviewed by the IRB of Indiana University Bloomington (``Gender
objectification in online conversations'', protocol \#1609276630).


\section{Data and Methods}\label{data_methods}

\subsection{Data and Terminology}\label{data} 

Our data comprises of all chat messages posted in public Twitch chat rooms
between August 26th and November 10th in 2014 (76 days). There were
1,275,396,751 messages posted in 927,247 channels (1,375 messages per channel
on average), by 6,716,014 viewers (190 messages per user on average). For each
message the following information is available: timestamp, author, channel, and
message text. Author and channel are identified by screen name; for the
channel, this is the screen name of the streamer.

\begin{figure}[t] 
\begin{center}
\includegraphics[width=\columnwidth]{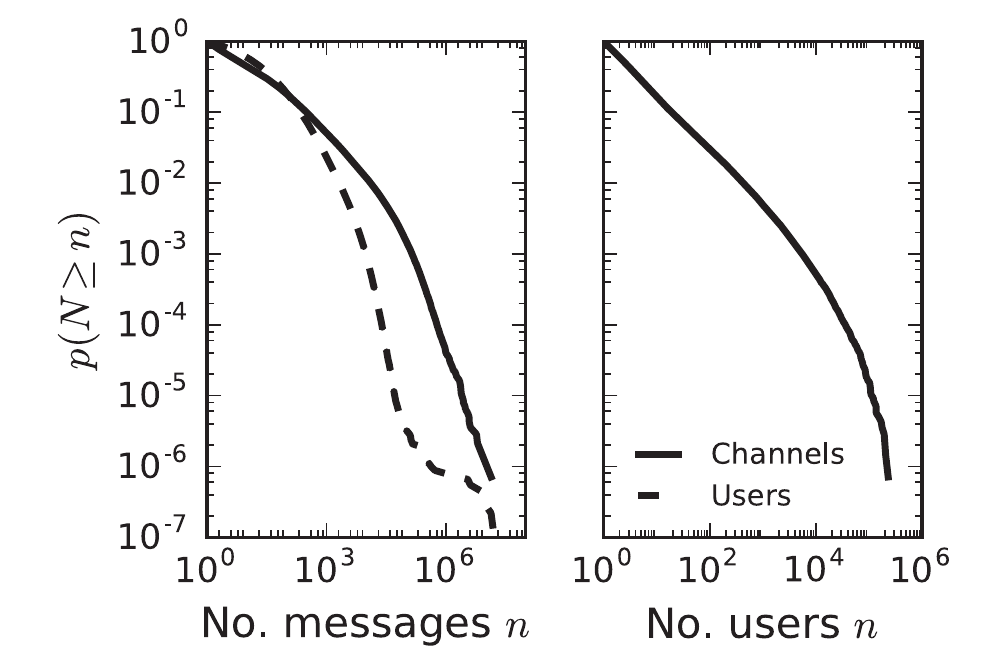}
\end{center}
\caption{Activity and popularity on the Twitch chat. \normalfont{Left: distribution of the
        number of messages produced by each channel (solid line) and by each user
        (dashed line). Right: distribution of the number of chatting users per
channel.}}
\label{fig:channel_activity_dist}
\end{figure}

Similar to other social media, the \emph{activity} and \emph{popularity} of
Twitch channels are highly skewed. We can quantify them by counting the number
of messages produced by each user and channel (for activity), as well as the
number of users chatting in a channel (for popularity).
Fig.~\ref{fig:channel_activity_dist} shows the distributions of these
variables.

Our analysis is based on the subset of 71,154,340 messages posted to the chats
of a matched sample of 200 female and 200 male streamers.  To estimate the
gender of streamers we manually examined the webcam feeds from archived video
feeds of past streams. We started by ranking all channels by the total number
of chat messages and examined the streamers of the most active 1,000 channels.
We discarded those streamers who do not share their webcam in their streams.
From this initial procedure we found 102 female streamers. From this group, we
discarded three streamers whose profile information was not written in English,
leaving 99 English-speaking female streamers from the top 1,000 streamers. We
then applied the same procedure to a random sample of less popular streamers,
i.e.\ whose channels ranked between $1,000$ and $16,000$ in the chat activity
distribution, and found the remaining 101 female streamers.

Having found a sample of female streamers, we identify a matched sample of male
streamers. The reason for matching samples is that prior work on these data has
shown that the nature of conversations on Twitch depends dramatically on chat
activity~\cite{nematzadeh2016twitch}. In particular, as the rate of messages
increases, messages become shorter and contain more emoticons. Because male
streamers tend to be on average more popular than female ones, and since more
popular channels will inevitably have a higher rate of chat activity,
statistical estimates of language difference may be biased.  To control for
this potential source of bias we thus match male and female streamers by
stratifying on the channel activity distribution~\cite{Rosenbaum2002}.

Because male streamers outnumber female ones, we sampled male streamers who
matched the 200 female streamers identified before. We used the number of chat
messages as the matching criteria; that is, every male streamer has a matching
counterpart in the female streamers sample with respect to the channel
activity, and not to the rank. As we did for the identification of female
streamers, the gender of male streamers was manually identified and only those
who used English in their profile were kept. For the remainder of this paper we
refer to the top 100 channels in each gender as \emph{popular} channels and the
rest as \emph{less popular} channels.

\subsection{Statistically Overrepresented Words}\label{subsection:log_odds} 

As our first exploratory analysis, we detect unigrams and bigrams that are
over-represented in either male or female streamers. To do so, we use log-odds
ratios with informative Dirichlet prior method~\cite{monroe2008fightin}. The
method estimates the log-odds ratio of each word \textit{w} between two corpora
\textit{i} and \textit{j} given the prior frequencies obtained from a
background corpus $\alpha$.  The log-odds ratio for word \textit{w},
$\delta_\textit{w}^{(\textit{i-j})}$ is estimated as
\begin{equation}
    \begin{split}
	\delta_\textit{w}^{(\textit{i-j})} & = \log\left(\frac{y^i_w +
\alpha_w}{n^i + \alpha_0- (y^i_w + \alpha_w)}\right) + \\ & -
\log\left(\frac{y^j_w + \alpha_w}{n^i + \alpha_0- (y^j_w + \alpha_w)}\right),
    \end{split}
\end{equation}
where $n^i$ (resp. $n^j$) is the size of corpus $i$ (resp. $j$), $y^i_w$ (resp.
$y^j_w$) is the count of word $w$ in corpus $i$ (resp. $j$), $\alpha_0$ is size
of the background corpus, and $\alpha_w$ is the frequency of word $w$ in the
background corpus.

Furthermore, this method provides an estimate for the variance of the log-odds
ratio,
\begin{equation}
\sigma^2(\delta_\textit{w}^{(\textit{i-j})}) \approx \frac{1}{(y^i_w +
\alpha_w)} + \frac{1}{(y^j_w + \alpha_w)},
\end{equation} and thus a $z$-score:
\begin{equation}
Z =
\frac{\delta_\textit{w}^{(\textit{i-j})}}{\sqrt{\sigma^2(\delta_\textit{w}^{(
\textit{i-j})})}}
\end{equation}
By leveraging the informative prior obtained from the background corpus, this
method often outperforms other methods such as PMI (point-wise mutual
information) or TF-IDF, particularly in detecting significant differences of
frequent words without over-emphasizing fluctuations of rare
words~\cite{monroe2008fightin, jurafsky2014narrative}.


\subsection{Word and Document Embeddings}\label{sub_section:doc2vec} 

To characterize individual users and channels we create a document for each
channel (and user) by aggregating every chat message in the channel (by the
user). Then we jointly obtain vector-space representations of words, users, and
channels by using the paragraph vector (\dv{}) method~\cite{le2014distributed},
which is an extension of the popular \wv{} embedding
method~\cite{mikolov2013efficient}.  This joint embedding not only allows us to
do vector operations across documents and words, but has been argued to
outperform other document embedding methods in document similarity comparison
tasks~\cite{dai2015document}. We use the skip-gram with negative sampling
(SGNS) model to learn word vectors. Among two main \dv{} models --- distributed
memory (DM) and distributed bag of words (DBOW) --- here we use the DBOW model
because of its conceptual simplicity, efficiency, and reported superiority in
performance~\cite{dai2015document}.  We use the \dv{} implementation available
in popular \texttt{gensim} Python library~\cite{rehurek_lrec}. The dimension of
vectors is set to 100 and the window (skip-gram) size is set to 5. The model is
trained with 10 epochs. All source code and the models used in this paper are
available on Github\footnote{https://github.com/\dots}






\section{Results}\label{results}

\subsection{Exploratory Language Analysis}\label{subsection:top_words_results}

We perform a term-based exploratory analysis to identify gendered terms. We
group channels based on their popularity (i.e.,~popular vs.~less popular) and
gender, producing four ``documents'': `popular male', `popular female',
`less popular male', and `less popular female'. To remove channel-specific
terms, we keep only words that are used in at least 100 channels, 20 female
channels, and 20 male channels. We make two comparisons: popular female
vs.~popular male and less popular female vs.~less popular male. We compute 
log-odds ratio (see \S~\ref{subsection:log_odds}) for unigrams and bigrams by
using the word frequency in the entire Twitch dataset as the prior.  The terms
are ranked by their estimated $z$-scores and the 25 terms with the largest
absolute $z$-score values are selected and visualized. The $z$-scores of the
displayed words range from 175 to 17. The identified words are then manually
categorized into four categories --- streamer ids, game-related jargon,
objectifying cues, and miscellaneous --- by using the information available in
Twitch and other online forums. By objectifying cues, we mean language that
reduce women to their body or appearance~\cite{langton2009sexual} or as objects
to be owned or used~\cite{nussbaum1995objectification}.

\begin{figure*}[t] 
  \begin{center}
      \includegraphics[width=\columnwidth]{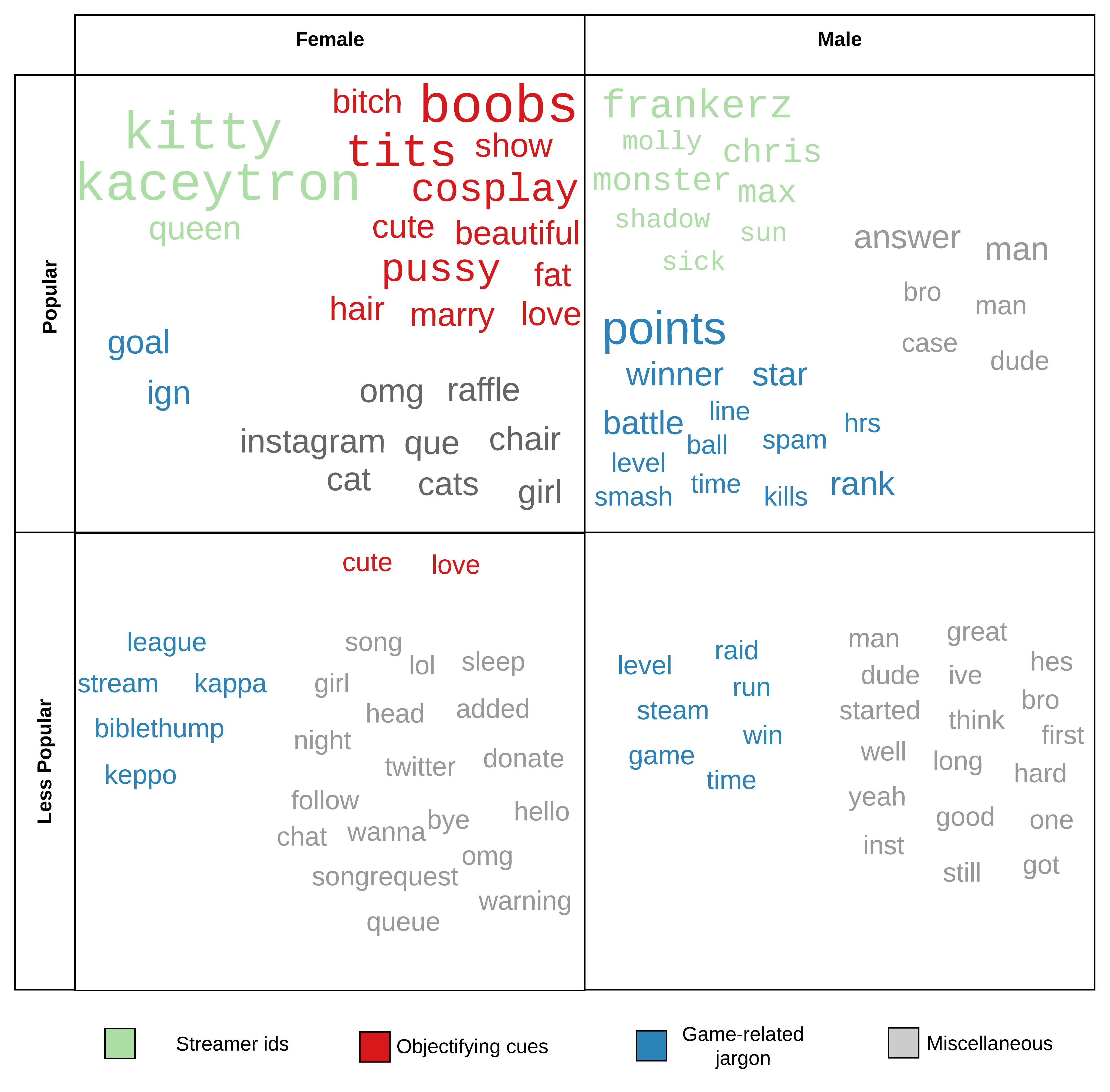}
      \includegraphics[width=\columnwidth]{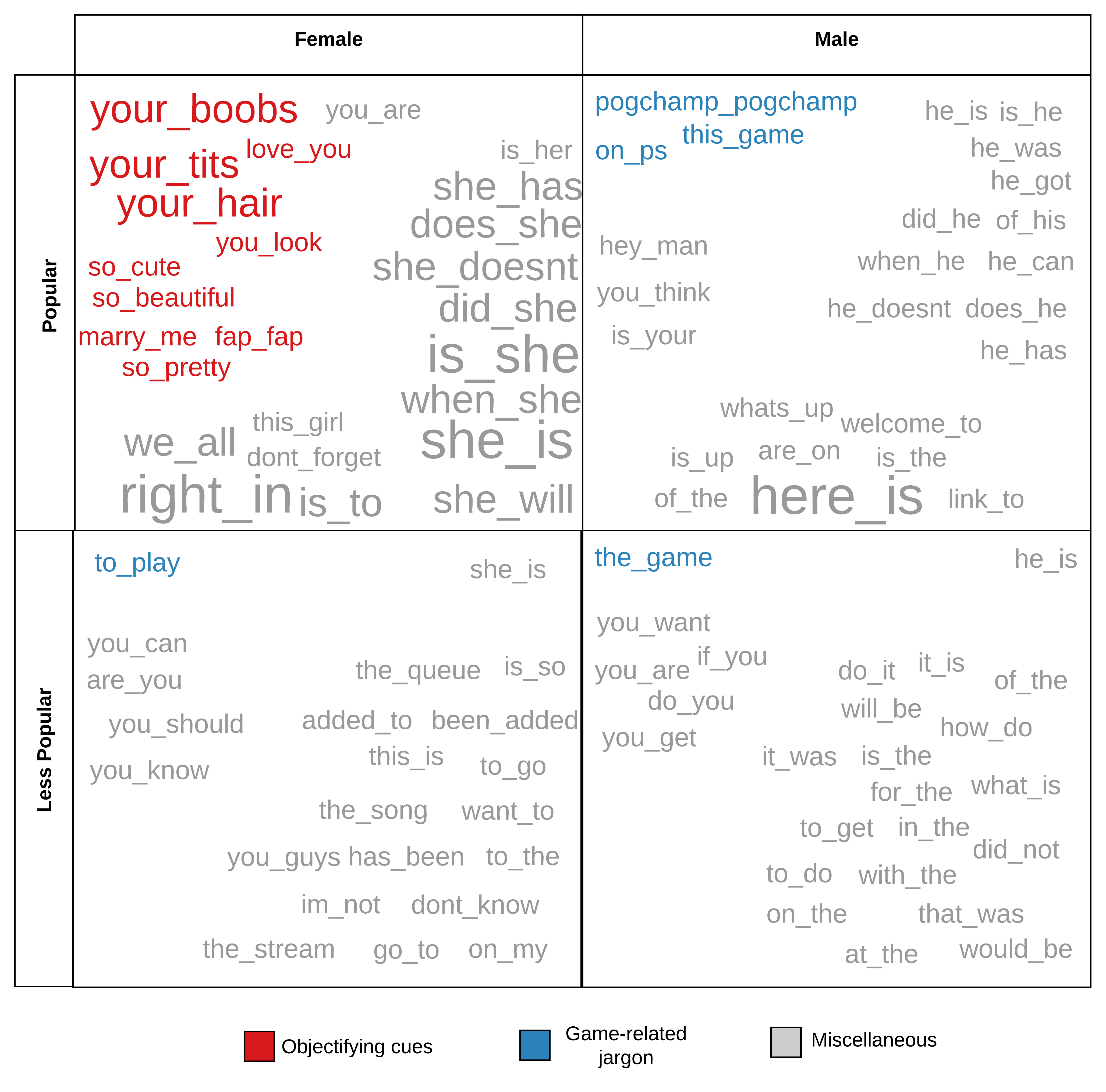}
  \end{center}

  \caption{Statistically over-represented $n$-grams in female and male channels
{\normalfont{} Left: unigrams. Right: bigrams. Font size is proportional to the $z$-score.}}
\label{fig:word_category_table}

\end{figure*}

Popular channels display a clear contrast between two genders
(Fig~\ref{fig:word_category_table}, left). Game-related words are clearly
overrepresented in male channels while words that signal objectification are
strongly associated with female channels. Interestingly, such words are not
apparent in less popular female channels. Instead, the words that signal
social interactions such as ``hello'', ``bye'', and ``song'' (i.e., automated
playlist requests) stand out in female channels. Also note the word
``warning''. This may suggest that the less popular female channels tend to
have stronger moderation in place, and tend to be used more as a social
gathering than a sporting event.

We repeat the same process for bigrams, which illustrate a similar pattern
(Fig.~\ref{fig:word_category_table}, right). Among popular channels, those belonging to
female streamers are characterized by terms, presumably directed at the
streamer, about their physical appearance; male channels are instead associated
with more game-related terms. Again, less popular female channels do not
show clear signs of objectification. Additionally, the results show the
contrasting use of second-person and third-person pronouns across popular and
less popular channels. In popular channels, even though objectifying cues seems
to be directed to the streamer, other bigrams are in third-person. By contrast,
less popular channels are dominated by second person pronouns. This observation
suggests that direct communication with the streamer dominates in less popular
channels while participating in popular channels resembles watching sporting
events.

\subsection{Analyzing Channels} 

To identify lexical features from female and male channels we train document
embedding models for the selected 400 channels. To identify non-trivial
gendered terms, we first remove trivial terms that signal gender: ``he'',
``she'', ``hes'', ``shes'', ``his'', ``her'', ``hims'', ``hers'', ``himself'',
``herself'', ``man'', ``woman'', ``bro'', ``boy'', ``sir'', ``dude'', ``girl''
and ``lady''. 

\begin{figure}[t] 
  \begin{center}
      \includegraphics[width=\columnwidth]{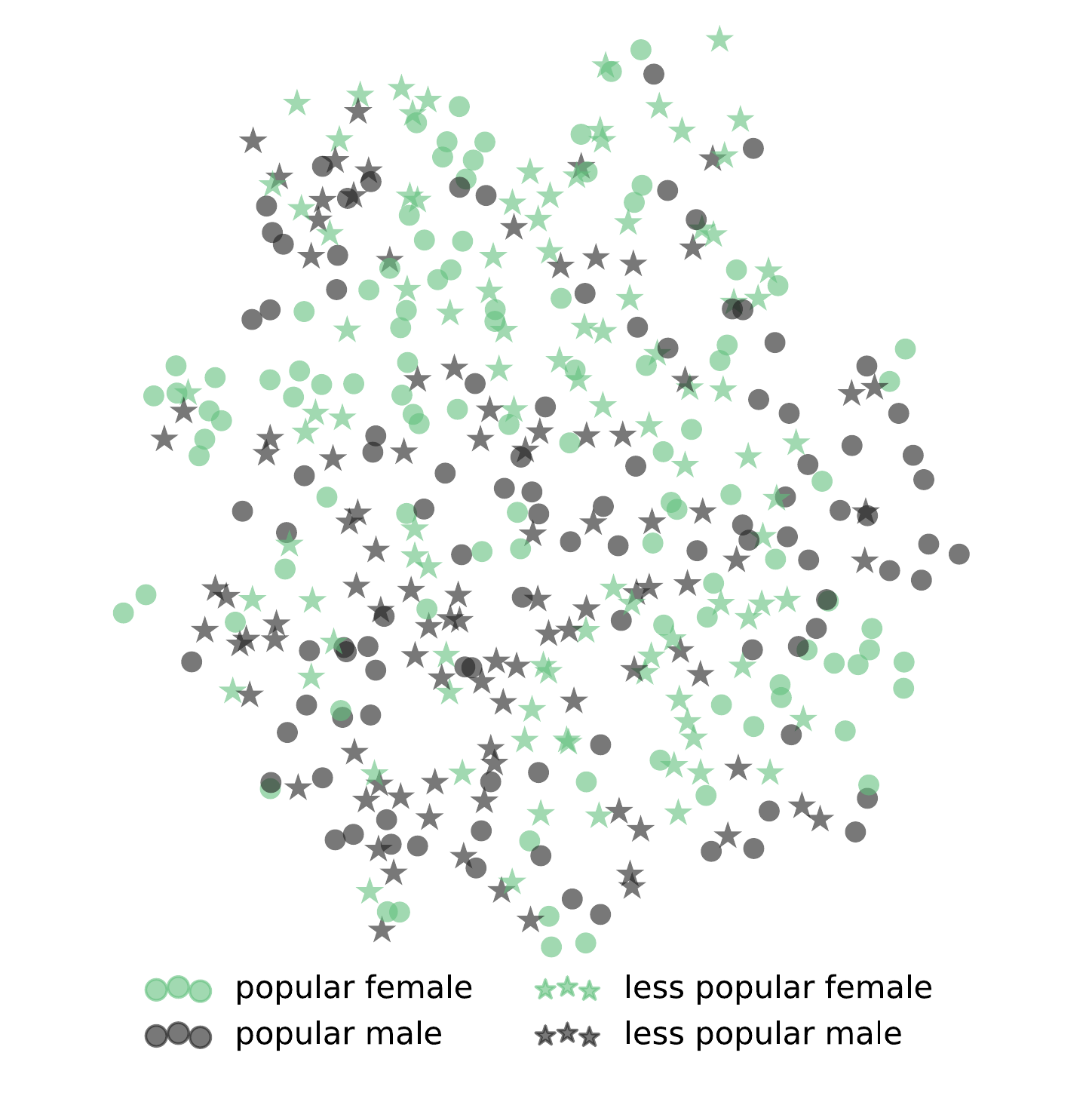}
  \end{center}

\caption{Visualization of channel vectors}\label{fig:channel_clusters_doc2vec}

\end{figure}

After the preprocessing we train the \dv{} model (see
Sec.~\ref{sub_section:doc2vec}). As noted, both document vectors and word
vectors are trained jointly. We visualize the vectors by applying \tSNE{}, a
popular manifold learning (dimensionality reduction)
method~\cite{maaten2008visualizing}. The map suggests clustered structure
based on gender (see Fig.~\ref{fig:channel_clusters_doc2vec}).

We then train two classifiers that predict the gender of a streamer by using L2
regularized logistic regression with two sets of features: bag of words (BoW)
and \dv{}. We evaluate the model by using 5-fold cross validation. As a
baseline, we use BoW model with 10,000 features using TF-IDF
vectorization~\cite{aggarwal2012mining}. This model exhibit the accuracy of
74\% ($\pm$~0.11\%, 95\% confidence interval) and a mean AUC of 0.80 in ROC curve
for the holdout test set. Then we use the normalized document vectors as the
features, obtaining an accuracy of 87\% ($\pm$~0.07\%, 95\% confidence interval)
and a mean area under curve of 0.93. 

Let us examine the key features. For the BoW model, we identify the words that
correspond to the largest absolute coefficient values (see
Table~\ref{table:channel_classification_features}). For the \dv{}-based model,
because it is not straightforward to connect each feature to a word, we use a
different approach; Since the \dv{} model learns documents and words vectors in
the same vector space, we simply identify the \emph{words} that are most
clearly identified as a female or male \emph{document}. We first extract the
top 10,000 words based on frequency in the channel corpus, then identify the
words that result in the highest (lowest) probability values in our
classification model (see Table~\ref{table:channel_classification_features}).

Our results indicate that female channels are characterized by words about
physical appearance, the body, relationships, and greetings while male channels
are characterized by game-related words. Male channels are also associated with
many uncommon words, suggesting that the male channel chats are more diverse
while the content in female channels share common words that signal
objectification. In sum, both our exploratory analysis and classification
exercise suggests that the answer to our first main research question --- ``are
the chat messages that streamers receive gendered?'' --- is ``yes''. 

\begin{table} 
  \centering
  \scalebox{0.8}{\begin{tabular}{p{1.5cm}p{4cm}p{4cm}}
    \toprule
      & \textbf{Doc2vec} & \textbf{BoW}\\
      \midrule
      \textbf{Female} & cute, beautiful, smile, babe, lovely, marry, boobs, 
      gorgeous, omg, hot &
          hi, boobs, song, hello, tess, emily, cat, love, cassie, kittens\\
      \midrule
      \textbf{Male} & epoch, attempts, consistent, reset, shields, fastest, 
      devs, slower, melee, glitch &
          frankerz, game, chris, got, adam, aviator, level, chief, arv, kynan\\
      \bottomrule
  \end{tabular}}
\caption{Channel classification learned features}
\label{table:channel_classification_features}
\end{table} 


\subsection{Analyzing Individual Users} 

Now we turn our attention from the \emph{streamers} and the chat messages that
they \emph{receive} to the \emph{viewers} and the messages that they
\emph{produce}. We ask whether the viewers and their chat messages are also
gendered, by examining gender preferences in channel selection by users, and their
linguistic differences.

\begin{figure}[t] 
  \begin{center}
      \includegraphics[width=\columnwidth]{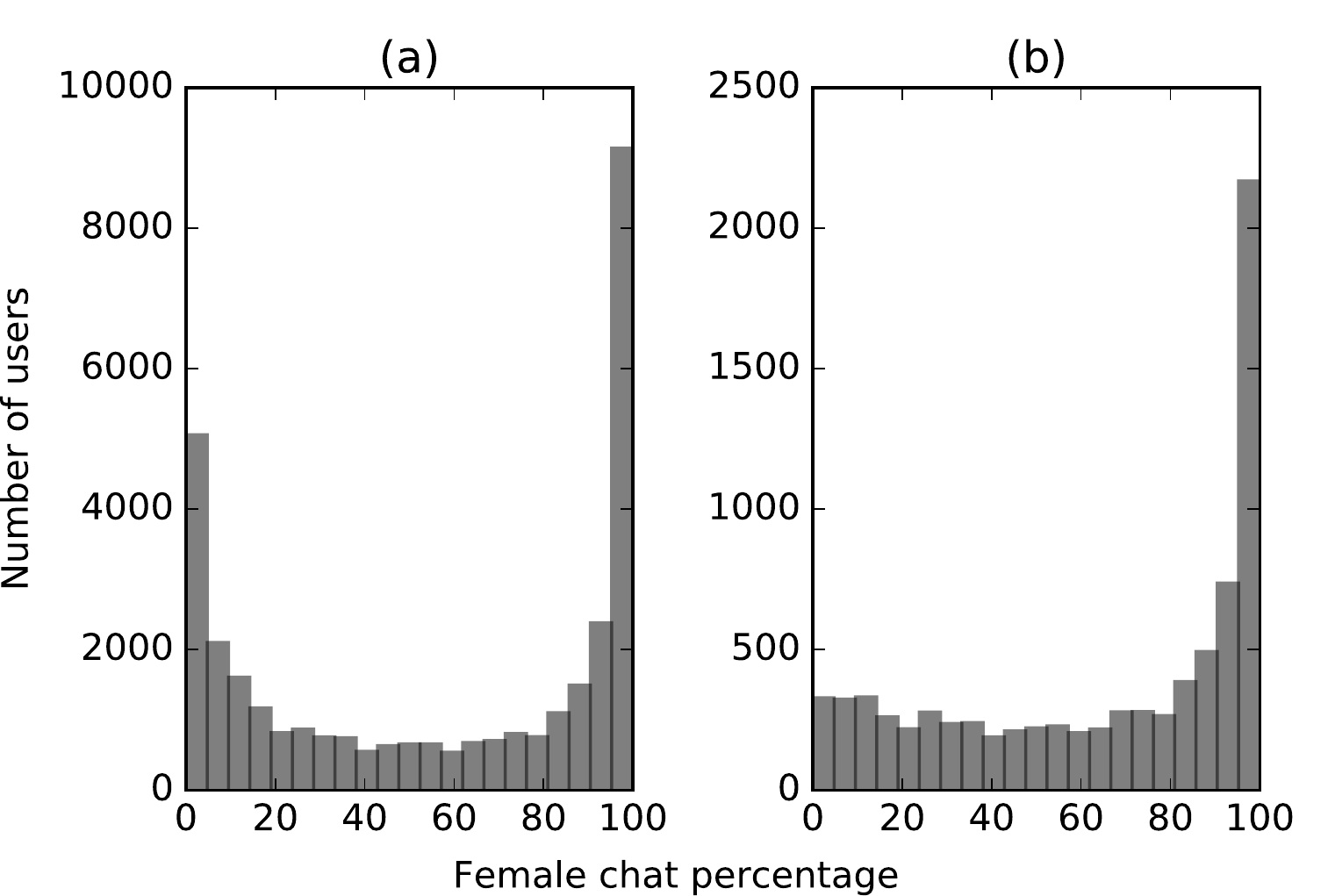}
  \end{center}

\caption{Gender preferences in channel selection. {\normalfont{} (a) Percentage of female
channels among the channels that a user posted (among users who posted in at
least 5 channel), (b) The same percentage among users who posted in at least 10
channels }}\label{fig:chat_percentages}

\end{figure} 

We examine whether the selection of which channels to post to is associated
with a given gender by calculating, for each user, the percentage of female channels
that they posted among all (400) channels. 1,818,028 individual users posted at
least one message in at least one of the 400 channels. We narrow it down to
93,898 (5\%) active users who have posted at least 100 messages, primarily to
obtain reliable language models. Our result suggests strong gender preference in channel
selection (see Fig.~\ref{fig:chat_percentages}). Note that if choices 
were random we would expect a binomial distribution with a peak at 50\%. By
contrast, we see a strong gender divide. A large fraction of
users (16\%), even when we focus on the users who have posted in more than five
channels, have posted only in male or female channels.  Moreover, if we limit
ourselves to users who have posted in many (10+) channels, a clear peak at
100\% --- \emph{10 female-channels and 0 male-channels} --- is visible,
indicating that the choice of chat participation is gendered and a significant
fraction (8\%) of users post messages only in female-streamer channels. Among
93,898 active users, there are 20,185 users and 21,883 users who posted only in
female or male channels respectively. Most of these users has posted messages
in popular channels~(see Table.~\ref{table:biased_user_distribution}).

\begin{table} 
  \centering
  \scalebox{0.8}{\begin{tabular}{p{4.6cm}p{1cm}p{0.8cm}p{1cm}p{0.8cm}}
    \toprule
      & \textbf{Female} & & \textbf{Male} &\\
      \midrule
      \textbf{Only Popular Channels} & 14,849 & (74\%) & 17,168 & (78\%)\\
      \midrule
      \textbf{Both} & 3,576 & (18\%) & 2,672 & (12\%)\\
      \midrule
      \textbf{Only Less Popular Channels} & 1,829 & (9\%) & 2,089 & (10\%)\\
      \midrule
      \textbf{Total} & 20,185 & (100\%) & 21,883 & (100\%)\\
      \bottomrule
  \end{tabular}}

\caption{Distribution of users who posted only in female or male channels}
\label{table:biased_user_distribution}

\end{table}

\begin{figure*}[t!] 
  \begin{center}
      \includegraphics[width=\textwidth]{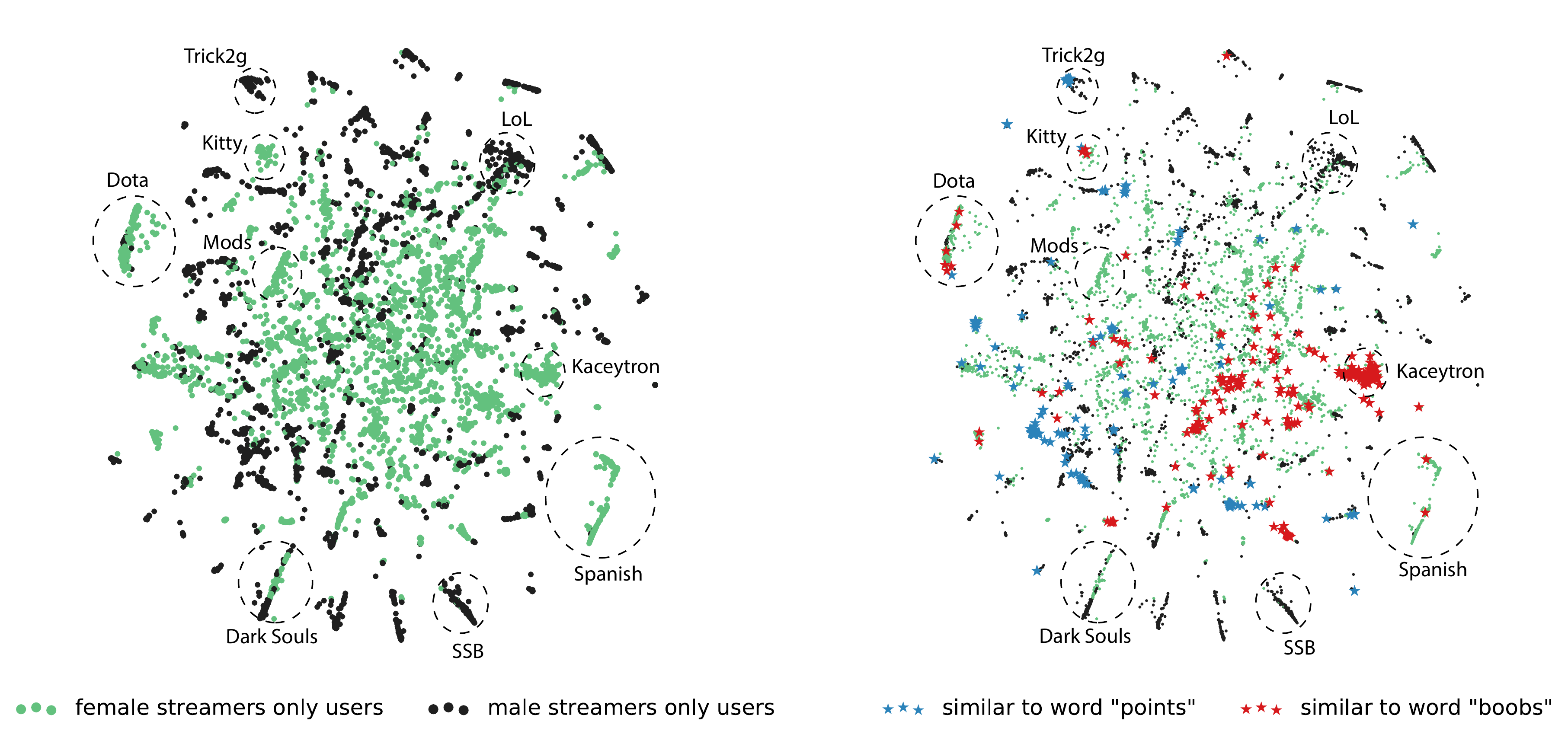}
  \end{center}

\caption{Visualization of user vectors using t-SNE}
\label{fig:user_clusters_word_sims}

\end{figure*}

Let us examine the linguistic differences between those strongly gendered
users. As described before, a user's chat messages is condered as a document
and its vector-space representation is obtained using \dv{} model. From 42,068
users who posted only in male or female channels, we randomly selected
10,000~(24\%) users (4,802 are female-only viewers) and examine them closely.
We first visualize their document vectors with \tSNE{}
(Fig.~\ref{fig:user_clusters_word_sims}, left). 

The map shows a clear separation between the two types of users, suggesting
clear lexical differences.  The map also shows distinct clusters. We study the
meaning of these clusters by identifying strongly associated words for each
cluster.  Specifically, given a cluster of $n$ document vectors $C = \{d_1,
d_2, \dots, d_n\}$, we find every word $w$ that is close to many of the
documents vectors, satisfying the following condition:
\begin{equation}
\frac{\bigl\lvert \{ d \in C | S_c(d, w) \ge s_{\min} \} \bigr\rvert}{|C|} \ge
f_{\min}, 
\end{equation}
where $S_c(d,w)$ is the cosine similarity between two vectors $d$ and $w$, and
$s_{\min}$ and $f_{\min}$ are two free parameters. We use $s_{\min} = 0.4$ and
$f_{\min} = 0.9$. 

We select eight clusters from the \tSNE{} map and identify characteristic words
for them (shown in Table~\ref{table:representatory_words}). The characteristic
words clearly signal the `topic' of the cluster, which we use as labels (see
Table~\ref{table:representatory_words} and
Fig.~\ref{fig:user_clusters_word_sims}). For instance, every representative
words in the ``League of Legends (LoL)'' cluster represents either a position
(``junglers'') or a character (``thresh'') in the game. Some of the identified
clusters are related to streamers (``Kaceytron'', ``Kitty'', ``Trick2g'') and
some are related to games (Dota, League of Legends (LoL), Super Smash Bros
(SSB), Dark Souls). We can also see two clusters related to Spanish language
and chat moderators. The terms in the ``Mods'' cluster are the ids of the users
who are known as moderators for multiple channels.  


\begin{table}[ht!] 
  \centering
  \scalebox{0.8}{\begin{tabular}{p{2cm}p{7.5cm}}
    \toprule
    \textbf{Cluster} & \textbf{Representative Words}\\
    \midrule
    Dota &
    mirana, slark, qop, potm, furion, slader, lycan, bristle\\
    \midrule
    LoL (League of Legends)  &
    champ, junglers, thresh, liss, azir, morg, riven, nid\\
    \midrule
    DarkSouls &
    freja, lucatiel, drangleic, artorias, darklurker, estus, smelter, dragon\\
    \midrule
    SSB (Super Smash Bros.) &
    ness, dedede, palutena, fow, wario, shulk, miis, jiggz\\
    \midrule
    Kaceytron &
    kaceytron, kacey, kaceys, catcam, kaceytrons, objectify, poopbutt, 
    objectifying\\
    \midrule
    Kitty &
    kitty, kittyplaysgames, moonwalk, kittys, kittythump, kittyapprove, 
    caturday, kittysub\\
    \midrule
    Trick2g &
    trick, godyr, dyr, dcane, trklata, trkcane, trkhype\\
    \midrule
    Mods (Moderators) &
    superfancyunicorn, tsagh, omeed, ironcore, tobes, snago, ara, moblord\\
    \midrule
    Spanish &
    dividir, jajajaja, palomas, carajo, belleza, negrito, aca, peruano\\
    \bottomrule
  \end{tabular}}

\caption{Representative words for the identified clusters using \dv{} and
\tSNE{}.}\label{table:representatory_words}

\end{table} 

To glean the relationships between these gendered users and their language, we
first pick the two most gender-biased words from
Sec.~\ref{subsection:top_words_results}: ``points'' and ``boobs''. We then
calculate the cosine similarity between each of the word and user document
vectors to identify the user document vectors that are most similar to one of
the two words. The top 250 users for each word are selected and overlaid on top
of the \tSNE{} map in Fig.~\ref{fig:user_clusters_word_sims} (Right).  The two
groups of users are clearly separated on the map.  Interestingly, the
identified user clusters tend to contain only one set of users. For instance,
``Kaceytron'' cluster is full of users whose vectors are highly similar to the
vector for ``boobs'', suggesting that the chat messages made by these users
share high semantic similarity with the word ``boobs''. Indeed, not only she
portrays a highly controversial streotype of female gamer (e.g.\ ``attracting
viewers with cleavage''), but also her channel is famous for not banning anyone
nor filtering any comments, and for directly responding to abusive
comments~\cite{kaceytron_interview}. By contrast, the ``Trick2g'' --- a famous
streamer for his game commentaries --- cluster only contains users who share
the context with ``points''. 

\begin{figure*}[htbp!]
  \begin{center}
      \includegraphics[width=\textwidth]{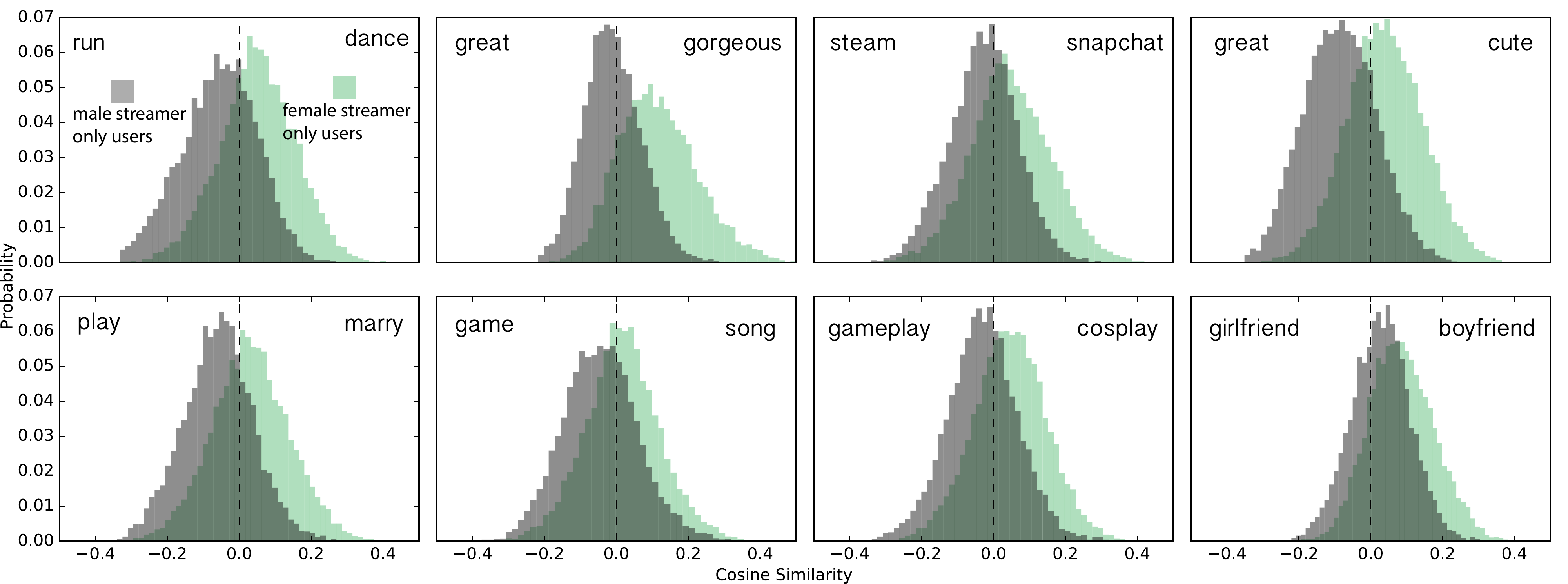}
      \caption{Cosine similarity skew}\label{fig:cosine_similarity_skew}
  \end{center}
\end{figure*}

Inspired by clear separability of users based on their gendered language, we
analyze how these gendered viewers are distributed with other gendered word
pairs. First, we manually selected eight gendered word pairs from our
exploratory analysis in Sec.~\ref{subsection:top_words_results}.  From these
pairs, we calculate the difference vector between the two word vectors; given a
pair of words, a game-related one $w_g$ and an objectifying one $w_o$, we
calculate $\vec{v}_{g \rightarrow o} = \overrightarrow{w_o} -
\overrightarrow{w_g}$. This vector, which roughly estimates the semantic
difference between those two vectors, is then used to project and compare each
user document vector. A positive cosine similarity value means the user vector
is closer to $\overrightarrow{w_o}$ and a negative value suggests the user
vector is closer to $\overrightarrow{w_g}$. The results, shown in
Fig.~\ref{fig:cosine_similarity_skew}, confirm our intuition.  Viewers who only
post in female-streamer channels tend to share similar context with words that
signal objectification, while those who post in male-stream channels only tend
to share similar context with game-related words.




Our analysis suggests that gendered viewers should be clearly separable based
on their language. So, we build classifiers; again, we train logistic
regression classifiers with BoW and \dv{} features.  In an evaluation using
5-fold cross-validation, BoW features achieve an accuracy of 96\% ($\pm$~0\%,
95\% confidence interval) and a mean area under curve of 0.99 in ROC curve,
while \dv{} features achieve 88\% ($\pm$~1\%, 95\% confidence interval) accuracy
and a mean area under curve of 0.95 in ROC curve.  The BoW model performed
surprisingly well in this classification task. However, our feature analysis
shows that it is a result of identifying trivial features.
Table.~\ref{table:user_classification_features} lists the most important
features in BoW model and most of them are channel-specific terms such as
streamer IDs. By contrast, the key \dv{} features, inferred by the method
described above, are more general gendered terms.

\begin{table} 
  \centering
  \scalebox{0.8}{\begin{tabular}{p{1.2cm}p{3.7cm}p{3.7cm}}
    \toprule
      & \textbf{Doc2vec} & \textbf{BoW}\\
      \midrule 
      \textbf{Female} & gorgeous, beautiful, makeup, wig, cute, marry, dress, 
      perv, pervert, smile &
          lea, kaceytron, cat, boobs, kitty, sheever, kacey, sonja, hafu, dizzy
          \\
      \midrule
      \textbf{Male} & ridley, quad, melee, cirno, glitch, unlocked, 
      leaderboards, mechanic, resets, rebirth &
          hutch, nelson, chris, boogie, warowl, fow, nickel, amp, aeik, moe\\
      \bottomrule
  \end{tabular}}

\caption{User classification learned
features}\label{table:user_classification_features}

\end{table} 

So far our main focus has been on the majority of users, who post only in male
or female channels. Next, in order to verify whether objectifying traits are
prevalent only in gender-biased individuals, we select a set of 2,734 \emph{balanced} users who
posted an approximately equal amount of messages in both female and male chat
rooms (female chat percentage between 40-60\%). We then separate those chat
messages into two groups based on the gender of the streamers and build
classification models to predict the gender of the streamer whose chat
the message was written in. Similar to previous analysis we train two classification models: one
using \dv{}, and another using BoW. The latter has an accuracy of 91\%
($\pm$~0.02, 95\% confidence interval) while the former 81\% ($\pm$~0.01, 95\%
confidence interval). Surprisingly, the learned features from both of these
methods are similar to the features learned from the analysis of users who post
only in a specific gendered channels (see
Table.~\ref{table:mixed_user_classification_features}), suggesting that
objectification is not a result of individual gender preferences, but is
commonplace.

\begin{table}[t] 
  \centering
  \scalebox{0.8}{\begin{tabular}{p{1.2cm}p{3.7cm}p{3.7cm}}
    \toprule
      & \textbf{Doc2vec} & \textbf{BoW}\\
      \midrule 
      \textbf{Female} & beautiful, cute, marry, cat, makeup, hair, cleavage, 
      hot, boyfriend, costume & kitty, boobs, lea, emily, tits, kaceytron, 
      ally, alisha, hafu, becca
          \\
      \midrule
      \textbf{Male} & bungie, gp, replay, hltv, jayce, blackscreen, comeback, vs, fp, chopper &
          moe, nelson, hutch, abdou, coty, chris, arnie, mr, boogie, bbg\\
      \bottomrule
  \end{tabular}}

\caption{Classification features of balanced users}
\label{table:mixed_user_classification_features}

\end{table} 

\section{Discussion}\label{discussion}

We reveal a nuanced picture of gendered conversation in Twitch, a social
game-streaming platform. Returning to the research questions we posed, our
analysis on both streamers and viewers shows that the \emph{conversation in
Twitch is strongly gendered}. First, the streamer's gender is significantly
associated with the types of messages that they receive --- male streamers
receive more game-related messages while female streamers receive more
objectifying messages. Second, the streamer's gender is also significantly
related to the channels viewers choose to watch. Many viewers choose to watch
and comment in only male or female channels and their messages are similarly
gendered; the messages posted by users who comment only in female channels tend
to have semantic similarity with objectifying cues while those who comment only
in male channels tend to have semantic similarity with more game-related terms.
Even the users who post in both male and female channels maintain similar
linguistic distinction based on gender; when posting to female channels they
tend to choose messages that have semantic similarity to objectifying cues.

Yet, we cannot unequivocally say that Twitch is a conversational hotbed for
gender stereotyping. In particular, the popularity of channels seems to change
the nature of chat, not only in terms of information
overload~\cite{nematzadeh2016twitch}, but also in terms of objectification,
moderation, and conversational structure.  Objectifying cues are only prevalent
in popular female channels. Less popular channels instead exhibit comments from
viewers that represent chat moderation.  Moreover, the user document embedding
technique reveals that there exist user clusters that consist of famous
moderators, indicating that strong, effective moderation is in place for many
less popular female-streamer channels.  Pronoun usage also changes depending on
the popularity of channels; less popular channels are characterized by usage of
second-person pronouns, signaling more intimate conversation \emph{between}
viewers and streamers, while popular channels exhibit the pattern that viewers
are talking \emph{about} streamers, except when they make objectifying remarks. 

We analyzed the language of users by employing multiple computational methods.
Our approach of using log-odds ratio with informative Dirichlet prior and the
\dv{} method effectively revealed the gendered nature of chat messages. \dv{}
allowed us to look at words and documents in the same space and also performed
better than the BoW features in document classification. \tSNE{} and vector
arithmetic proved useful in identifying clusters of terms and users. Our
methods contribute to growing literature on constructing language models to
identify and unpack gendered phenomena; for instance, we can draw parallel to
models by Fu et al.~\cite{fu2016tie} that found questions posed by journalists
to professional female tennis players objectified women, while questions posed
to male players focused were game-related, and by Way et
al.~\cite{way2016gender}, who found subtle gender inequalities in faculty
hiring practices among universities of different rankings and career
trajectories.

We acknowledge that our analysis has several limitations. Most notably, we
provide only a \emph{static} picture of Twitch limited to questions about
association rather than causal relationships. While we can only surmise the
causes of the gendered conversation we observed, financial motivations may
commodify and incentivize the objectification of female streamers.  Twitch
provides revenue for streamers through a subscription system, and many
streamers also deploy donation systems for additional revenue. Thus, financial
incentives exist for streamers to increase subscribers and possibly to conform
to the requests of the male viewers, the majority of many streamers'
``customers''. Such incentives may solidify the popularity of female streamers
who do not address (or even encourage) objectification, facilitating abusive
behavior against female gamers. This vicious cycle may reinforce and spawn the
structural problem of gender imbalance in online social gaming communities. If
part of feminism's remit~\cite{butler2011gender} is to consider how society,
females included, may play a role in constructing what a legitimate female's
identity is in, for example, online spaces, we argue that we should investigate
how Twitch supports heteronormative stereotypes. Future work may also examine
how pathways to popularity differ for male and female channels. For instance,
do female-streamer channels gradually evolve to conform to gender stereotypes
or allow objectifying comments? 

Our study also does not investigate how streamers themselves engage viewers
and the chat. There are a wide range of streamers from those who play games
without talking or chatting to those who actively engage with viewers
through frequent gaming events for their audience. Analyzing streamer behavior
is a challenging task requiring analysis of both the audio and video feeds of
streamers; emerging techniques for analyzing multimedia data may facilitate
future work examining the interplay between streamer behaviors and viewer
behaviors.


Last but not least, our work points to the need to examine the vast number of
small communities, albeit not so popular, on Twitch whose conversations do not
follow gender lines. Our analysis shows the existence of vigilant user groups
who provide moderation services to ensure the conversations revolve around
game-related topics.  This observation paints a less bleak picture of social
gaming. Might there be a way to bridge between these two disparate spaces of
crowded and intimate spaces?  Developing methods for automatic detection of
abusive and objectifying comments as well as other scalable communication and
moderation techniques will also be beneficial for online gaming communities.

\subsection*{Acknowledgements}
We 
thank Cyrus Hall, Spencer Nelson, and Drew Harry for providing
access to the Twitch chat log data. 

\printbibliography

\end{document}